\documentclass[12pt,preprint]{aastex}

\usepackage{graphicx}

\newcommand{\HI}{{\sc H\,i}}

\shorttitle{Radio Polarimetry of ELAIS N1}
\shortauthors{Taylor et.\ al.}

\begin{document}

\title{Radio Polarimetry of the ELAIS N1 Field:  \\
Polarized Compact Sources}

\author{A.\ R.\ Taylor, J.\ M.\ Stil, and J.\ K.\ Grant}
\affil{Department of Physics and Astronomy, University of Calgary}

\author{T.\ L.\ Landecker, R.\ Kothes, R.\ I.\  Reid, A.\ D.\ Gray }
\affil{Dominion Radio Astrophysical Observatory, Herzberg Institute of Astrophysics \\
National Research Council of Canada}

\author{Douglas  Scott}
\affil{Department of Physics and Astronomy, University of British Columbia}

\author{P.\ G.\ Martin, A.\ I.\ Boothroyd}
\affil{Department of Astronomy, University of Toronto}

\author{G.\ Joncas}
\affil{D\'epartement de physique, g\'enie physique et d'optique, Universit\'e Laval}

\author{Felix\ J.\ Lockman}
\affil{National Radio Astronomy Observatory}

\author{J.\ English}
\affil{Department of Physics, University of Manitoba}

\author{A.\ Sajina}
\affil{Spitzer Science Center, California Institute of Technology}

\and

\author{J.\ R.\ Bond}
\affil{Canadian Institute for Theoretical Astrophysics, University of Toronto}

\begin{abstract}
We present deep polarimetric observations at 1420 MHz of the European
Large Area {\sl ISO} Survey North 1 region (ELAIS N1) as part of the
Dominion Radio Astrophysical Observatory {\sl Planck} Deep Fields
project.  By combining closely spaced apertures synthesis fields, we
image a region of 7.43 square degrees to a maximum sensitivity in
Stokes $Q$ and $U$ of $78\,\mu$Jy beam$^{-1}$, and detect 786 compact
sources in Stokes $I$.  Of these, 83 exhibit polarized emission.  We
find that the differential source counts ($\log N - \log p$) for
polarized sources are nearly constant down to $p > 500\,\mu$Jy, and
that these faint polarized radio sources are more highly polarized
than the strong source population. The median fractional polarization
is $4.8 \pm 0.7$\% for polarized sources with Stokes $I$ flux density
between 1 and 30 mJy; approximately three times larger than sources
with $I > 100$ mJy.  The majority of the polarized sources have been
identified with galaxies in the {\sl Spitzer} Wide Area Infrared
Extragalactic Survey (SWIRE) image of ELAIS N1.  Most of the galaxies
occupy regions in the IRAC 5.8/3.6 $\mu$m vs.\ $8.0/4.5\,\mu$m
color-color diagram associated with dusty AGNs, or with ellipticals
with an aging stellar population.  A few host galaxies have colors
that suggests significant PAH emission in the near-infrared.  A small
fraction, 12\%, of the polarized sources are not detected in the SWIRE
data.  None of the polarized sources in our sample appears to be
associated with an actively star-forming galaxy.
\end{abstract}


\section{Introduction}

Observation of polarized radiation at radio wavelengths is one of the
prime means to study the roles of magnetic fields in astrophysics,
through synchrotron emissivity which samples magnetic fields in
relativistic plasmas, and through Faraday Rotation, created by
radiation propagation through magnetized thermal plasmas.
Understanding the origin and evolution of magnetic fields is a key
science goal of the The Square Kilometre Array, a next-generation
radio telescope under development by the international community
\citep{schilizzi2004}.  The primary observational tool to study the
magnetic universe will be an SKA all-sky Rotation Measure survey of
background radio sources and diffuse Galactic emission down to
polarized flux density levels of $\sim0.1\,\mu$Jy \citep{bg04}.
However, while modern source counts approach flux density
sensitivities of $\sim10\,\mu$Jy in total intensity
\citep{windhorst2003,hopkins2003}, very little is known about the
polarization properties of the faint radio source population.  The
most extensive analyses of polarization of compact extragalactic
sources were carried out by \citet{mesaetal2002} and
\citet{tucci2004}, who used the NVSS data \citep{condon1998} to derive
statistical polarization properties for $\sim$30,000 sources with
$S_{\rm 1.4GHz} > 100$ mJy.  Similarly, \citet{bg04} used NVSS sources
with total flux density greater than 80 mJy to extrapolate polarized
source counts to $\mu$Jy levels.

\citet{mesaetal2002} found that the mean fractional polarization of
radio sources in the NVSS brighter than 80 mJy was anti-correlated
with flux density, especially for steep-spectrum radio sources
($\alpha < -0.5$, for $S_\nu \sim \nu^\alpha$). \cite{tucci2004}
confirmed this result for the median of the fractional polarization
for steep-spectrum sources only (87\% of their sample), but found no
significant trend for flat spectrum sources (13\% of their sample).
\cite{tucci2004} also noted that the flat shape of the polarized
source counts indicates a dependence of the fractional polarization on
flux density.

Radio sources with $S_{\rm 1.4GHz} > 100$ mJy are predominantly
associated with flat or steep spectrum Active Galactic Nuclei. Star
forming galaxies begin to be a significant fraction of the population
at flux densities less than a few mJy (see e.g. Hopkins 2000,
Windhorst 2003). However, there is still ongoing debate about the
fraction of the sub-mJy radio sources which is radio-quiet active
galactic nuclei (AGN) \citep{gruppioni1999,simpson2006}. It is thus
uncertain that polarization properties derived from the strong radio
source population are applicable to the sub-mJy radio sources.

We have begun a project called the DRAO Planck Deep Fields to explore
the high latitude sky at high sensitivity in polarized radio continuum
and in atomic hydrogen emission.  The project uses the DRAO Synthesis
Telescope at 1.4 GHz to create deep images of two fields, one with a
very low column of foreground material, the ELAIS N1 region $(l,b) =
(84\degr, +45\degr)$, and a larger region of highly structured
infrared cirrus emission at $(l,b) = (135\degr, +40\degr)$. This paper
reports initial results from the first 30\% of observations of ELAIS
N1.  The ELAIS N1 (European Large Area {\sl ISO} Survey North 1) field
\citep{oliver2000}, is an area of approximately 2 square degree chosen
for a mid-infrared survey of distant galaxies with {\sl ISO}. The
field was selected to minimize confusion with Galactic cirrus and the
zodiacal background, being one of the lines of sight to the
extragalactic sky with minimum {\sl IRAS} 100 $\mu$m emission.  A
larger area that includes ELAIS N1 was later observed by the {\sl
Spitzer} Wide Area Infrared Extragalactic Survey (SWIRE)
\citep{lonsdale03}.  The large amount of archival data from these
extragalactic surveys make this region ideal for studies of the faint
polarized radio source population.

\section{Observations and Data Processing}
\subsection{Synthesis Observations}

The DRAO Synthesis Telescope (DRAO ST) is described in detail in
\citet{landecker2000}.  The telescope is a seven-element east-west
array of 9-m diameter antennas.  Three antennas are moved to
provide complete sampling of the UV plane from the shortest baseline
(12.86 m) to the longest baseline (617.18 m) after a full synthesis of
12 times 12 hours.  The array has a primary beam
size 107\farcm2 (FWHM) at $1420\,\rm MHz$, which makes it an effective
instrument for wide-field surveys. The first sidelobe of the
synthesized beam is at the 3\% level, and side-lobes farther from the main
lobe of the beam are less than 0.5\%. The first grating ring of the
synthesized beam appears at $2\fdg8$ from the main lobe at 1420 MHz,
which is outside the field of view defined by the 10\% sensitivity
level of the primary beam.

The telescope observes simultaneously the \HI\ 21 cm line and
continuum at 408 MHz and full polarimetry in four 7.5 MHz wide
frequency bands centered around $1420\,\rm MHz$.  The antennas have
prime focus feeds and at 1420 MHz receive both right-hand (R) and
left-hand (L) circularly polarized radiation.  The observations and
data processing techniques used in this paper follow those that have
been used to obtain high-fidelity wide-field polarimetric images of
the Galactic plane with the DRAO ST \citep{taylor2003,landecker2007}.

The system temperature of the telescope as described by
\citet{landecker2000} was $60\,\rm K$, leading to an rms noise in a
synthesized image of a single field of $280\,\mu$Jy~beam$^{-1}$ at
field center. However, starting in 2003, the sensitivity was enhanced
by a series of improvements, completed by the time the current
observations began in 2004. These improvements comprised installation
of new low-noise amplifiers, modifications to telescope structures to
reduce ground noise, and installation of shielding fences to further
block ground radiation from entering the aperture. The system
temperature of the seven individual antennas now spans the range
$35\,\rm K$ to $60\,\rm K$, and the overall system temperature is
$\sim45\,\rm K$, leading to a field-center rms noise of
$210\,\mu$Jy~beam$^{-1}$ ($53 \sin \delta\ \rm mK$) at 1420 MHz.  A
higher sensitivity and larger field of view are obtained by creating a
mosaic of overlapping fields. The most uniform sensitivity across a
mosaic is obtained if the fields are centered on a hexagonal grid. The
field centre separations for the survey presented here is $22\arcmin$,
which is $20.5 \%$ of the FWHM diameter of the primary beam.

\begin{figure}[h]
\begin{center}
\resizebox{14cm}{!}{\includegraphics[angle=0]{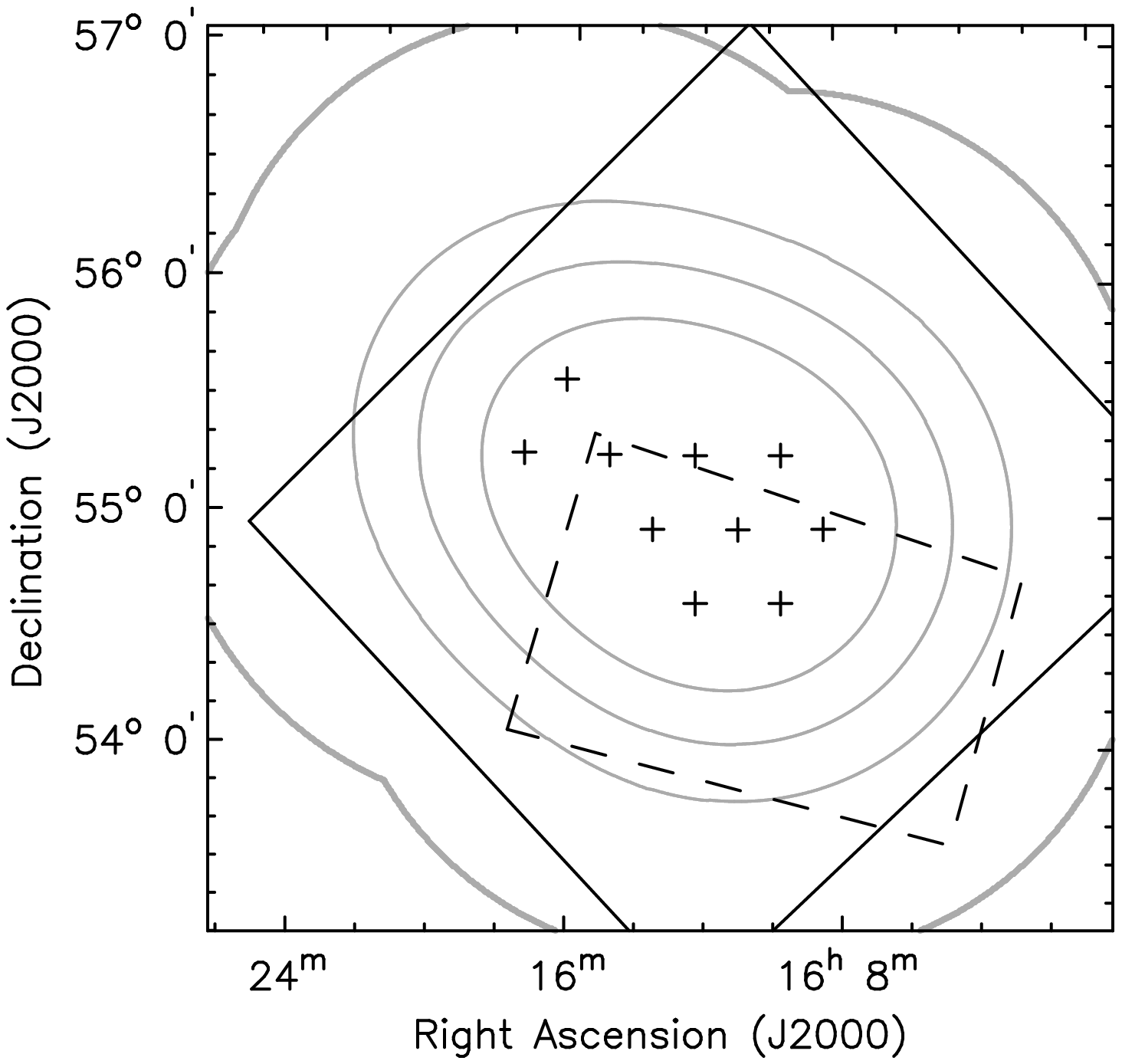}}
\caption{ Survey area of the first 10 fields of the DRAO mosaic in
relation to the sky coverage of the original ELAIS N1 field (dashed
lines) and the SWIRE survey (solid black lines).  The ten field
centers are indicated by the $+$ symbols. Gray contours indicate the
sensitivity at 1.5, 2, and 3 times the theoretical noise level in the
center of the mosaic, and the edge of the field of view (thick
line). 
\label{surveyarea-fig}
}  
\end{center}
\end{figure}

Observations
for the DRAO ELAIS N1 survey began in August 2004 and will continue to
create a final mosaic of 30 fields.  This paper presents 21-cm
continuum polarimetry of the first 10 fields.
Figure~\ref{surveyarea-fig} shows the location of the 10-field mosaic
in relation to the {\sl ISO} ELAIS N1 field, and the SWIRE survey of ELAIS~N1. 
The theoretical maximum 1-$\sigma$ sensitivity in our mosaic of 10 fields is $\sim80\,\mu$Jy.

\subsection{Polarization Data Processing and Calibration}

Complex gains for the center of each field were determined by 
observing the unresolved and unpolarized sources 
3C\,147 and 3C\,295 between 12-hour observing runs. The absolute
polarization angle was calibrated by observing the highly 
linearly polarized source 3C\,286 once every 4 days. 

The polarization images do not contain much flux, making
self-calibration ineffective. Therefore R and L gain solutions derived from
self-calibration of the Stokes $I$ images were applied to the polarization
data as well. The resulting visibility data sets for each field were
then corrected for the effects of instrumental polarization across the
field of view (which leads to leakage of Stokes $I$ power into $Q$ and
$U$). Instrumental polarization is a complex quantity, with phase and
amplitude terms.  These were measured for the seven individual antennas
with a holographic technique, using the unpolarized source 3C~295, on a
$15'$ grid across the primary beam. After interpolation, these
measurements were used to predict conversion of $I$ into $Q$ and $U$ at
any point in the beam based on CLEAN component models from processing
the $I$ image for a field. Residual errors in instrumental polarization
for an individual field are estimated at 0.25\% for the field center,
growing to 1\% at a distance of $75'$. With a field separation in the
mosaic image of $22'$ the instrumental polarization in the central 3.6
square degrees of the mosaic remains less than 0.5\%. Towards the edge
of the mosaic instrumental terms may be as large as 1\%.

After these initial procedures there are usually still confusing
arc-like structures left in the images, centered on bright sources
inside and outside the primary beam. These are the result of residual
complex gain errors at large distance from the field center. Effects
from these are removed from the visibilities using a procedure called
modcal, which is in principle a direction-dependent self-calibration
\citep{willis99}.

The antenna sidelobes are highly polarized, and sources outside the
primary beam can produce strong spurious polarized signals. The Sun is
seen in the sidelobes whenever it is above the horizon, but its effects
are usually confined to the shorter baselines because of its large
extent. The effects of the Sun were removed by making images centered on
the Sun's position and removing the response from the visibilities.
Terrestrial interference, which is always polarized, is another source
of spurious polarization. Interference effects in images appear as a
spurious source concentrated around the North Celestial Pole, and can be
largely eliminated by making an image at the Pole and correcting the
visibilities.  Radiation from the ground appears to be polarized, and
can appear as a correlated signal in visibilities corresponding to short
baselines; this effect is more difficult to remove. In some particularly
bad cases the data for affected interferometer spacings were simply
flagged and removed.

\subsection{The Images}

Figure~\ref{EN21-fig} shows the deep 21 cm continuum images of the
ELAIS N1 area in Stokes $I$, $Q$, and $U$.  The images are centered on
$\alpha_{2000}$ = $16^{\rm h} 11^{\rm m} $, $\delta_{2000}$ =
$+55\degr$ and cover an area of 7.43 square degrees (the area within
the thick gray line in Figure \ref{surveyarea-fig}).  The angular
resolution varies with the declination over the images and is given by
$b_\alpha \times b_\delta = 49'' \times 49'' {\rm cosec } \delta $.
At the mosaic center the resolution is $49'' \times 59''$.  The noise
near the center of the mosaic (white square in Figure~\ref{EN21-fig})
is measured to be $78\,\mu$Jy in $Q$, and $U$ (see
\S~\ref{detect-sec}). The Stokes $I$ image is not limited by confusion,
but the noise is slightly higher at $85\,\mu$Jy near the center of the
mosaic, probably because of a contribution from faint sources.

The final images are virtually free of artifacts, so the sensitivity is
limited by the noise.  The rms brightness sensitivity is $17.4\,\rm mK$. 
The dynamic range near the center of the mosaic is more than
3000:1 in Stokes $I$. With very few exceptions, sources in
Figure~\ref{EN21-fig} appear as compact (nearly) unresolved sources,
that can be characterized by their peak intensity and a single
polarization angle. This is consistent with expectations from the
angular size -- flux density relation for extragalactic radio sources
\citep{windhorst2003}; the median angular size of a $1\,\rm Jy$ source is 
$\sim 10''$.

\section{Compact Polarized Sources}

\subsection{Source Detection}
\label{detect-sec}

Flux densities and positions of all sources in the pilot deep field
images were measured with a source extraction algorithm that fits a
two-dimensional Gaussian and a background level to each source.  The
mosaic images were multiplied by the primary beam response function of
the mosaic to obtain an image with a uniform noise level, equal to the
noise level at the center of the image. This operation retains the
correct signal-to-noise ratio for each source, but the resulting
uniform noise level greatly facilitates automated source extraction.
The inverse primary beam correction is applied to the measured flux
densities of sources from the uniform noise map to transform back to
true flux density.  The rms noise level in the uniform noise images
was measured by fitting a Gaussian function to the distribution of
amplitudes in the image.  The result is shown in
Figure~\ref{fig:Q_Unoise}.  The noise distribution in both the $Q$ and
$U$ images is well fitted by a Gaussian with dispersion $\sigma =
78\,\mu$Jy.

The polarized flux density image ($p = \sqrt {Q^2 + U^2}$) was searched
for polarized sources.  For Gaussian noise in the $Q$ and $U$ images with
rms amplitude $\sigma$, the statistical distribution of the noise in
a measurement of $p$ for a source with an intrinsic polarized flux
density of $p_o$ is a Rice distribution
\citep{rice1945,vinokur1965,simmons1985}
\begin{equation}
f(p|p_{\rm o}) = \frac{p}{\sigma} e^{-\frac{(p^2 + p_{\rm o}^2)}{2 \sigma^2}} 
I_{\rm o}(\frac{p p_{\rm o}}{\sigma^2}) \ ,
\end{equation}
Here $I_{\rm o}$ is
the modified Bessel function of the first kind.  For $p_{\rm o} = 0$, the
distribution reduces to a Rayleigh distribution,
\begin{equation}
f(p|0) = \frac{p}{\sigma} e^{-\frac{p^2 }{2 \sigma^2}} \ ,
\end{equation}
which gives the probability distribution of pixel amplitudes in the
$p$ image in the absence of polarized emission.  The noise in the $p$
image has a non-zero mean and has higher probability of positive peaks
above a given detection threshold than Gaussian noise.  We searched
the $p$ map down to a level of $4.55\sigma$, which has an equivalent
probability for false positive signals to the $4\sigma$ level for a
Gaussian distribution.  The measured polarized flux density $p$ was
corrected for noise bias to obtain an estimate of $p_{\rm o}$ through
the relation $p_{\rm o}^2 = p^2 - \sigma^2$, which is a good
approximation if the signal to noise ratio is larger than 4
\citep{simmons1985}.

The 83 sources detected are listed in Table~\ref{table_polsources},
which gives the position of each source, the integrated total flux
density, noise bias-corrected peak polarized intensity, polarization
position angle, fractional polarization defined as the ratio of the
bias-corrected peak polarized intensity to the peak total intensity,
and the spectral index of the total flux density between $325\,\rm
MHz$ and $1420\,\rm MHz$ if the source appears in the WENSS catalogue
\citep{rengelink1997}.  To ensure possible spurious polarized sources
due to instrumental polarization are not included, we conservatively
remove sources with observed fractional polarization, $\Pi$, less than
1\%. Only two sources were removed from the sample for this reason.

\subsection{Distribution of Fractional Polarization}
\label{maxlike-sec}

\begin{figure}
\begin{center}
\resizebox{14cm}{!}{\includegraphics[angle=0]{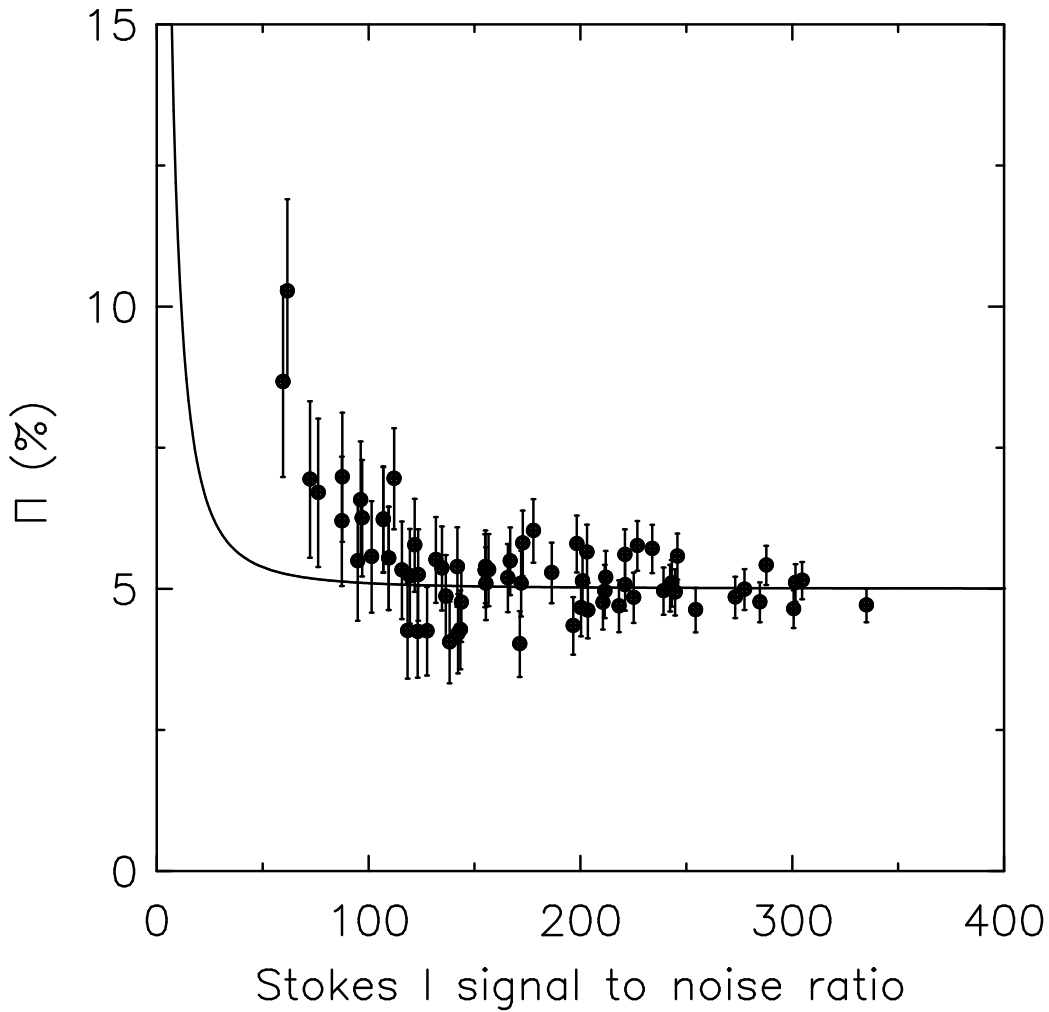}}
\caption{ Illustration of the effect of noise and the polarized flux density
detection threshold on $\Pi$ as a function of flux density. This figure shows
the variation of $\Pi$ with flux density for a simulated sample of
sources, all 5\% polarized, with random Gaussian noise added in $I$, $Q$,
and $U$.  The error bars are derived from standard error propagation,
assuming the noise is known. The curve shows the effect of polarization
bias on $\Pi$, defined as $\sqrt{(p_{\rm o}^2+\sigma^2)}/I$ \citep{simmons1985}. 
\label{polstat_demo}
}  
\end{center}
\end{figure}

The intrinsic fractional polarization $\Pi_{\rm o}$ of radio sources
provides astrophysical information about the nature of the polarized
sources.  However, the observed fractional polarization $\Pi$ is
sensitive to the noise in $p$ and in $I$. In addition to $p$ being a
biased estimator of $p_{\rm o}$, the error distribution of the ratio
$p/I$ has strong non-Gaussian wings, so $\Pi$ is not a very accurate
estimate of $\Pi_{\rm o}$ even for relatively high signal-to-noise
sources.  This is illustrated in Figure~\ref{polstat_demo}, which
shows the relation between $\Pi$ and Stokes $I$ flux density for an
artificial sample of sources, all with $\Pi_{\rm o} = 5$\% and
Gaussian noise added with equal amplitudes in Stokes $I$, $Q$, and
$U$.  Only those sources with a $p$ flux density more than $5\sigma$
are shown. Error bars represent standard error propagation in $\Pi$
from the errors in $p$ and $I$. The high values of $\Pi$ at
signal-to-noise ratio less than $\sim$100 is a result of the detection
threshold in $p$ and the non-Gaussian error statistics of $\Pi$. The
effect is much larger than the polarization noise bias alone.

These problems highlight the need for careful analysis of the effects
of noise and polarization detection thresholds in studies of the
fractional polarization of faint sources.  Previous studies have
focused on the fractional polarization of bright radio sources in the
NVSS. \citet{mesaetal2002}, \citet{tucci2004}, and \citet{bg04}
considered polarized sources in the NVSS with Stokes $I$ flux density
stronger than 80 mJy, 100 mJy, and 80 mJy respectively.  The high flux
density thresholds in these studies were adopted to achieve a good
level of completeness in $\Pi$ down to the limit set by residual
instrumental polarization ($\Pi \approx 1\%$). These studies found a
$\Pi$ distribution that decreases monotonically with increasing $\Pi$,
and with a median $\Pi \approx 1.8\%$.  For these bright sources,
noise effects are small, and the $\Pi$ distribution should be close to
the intrinsic $\Pi_{\rm o}$ distribution down to limits set by
residual instrumental polarization.

In this paper we investigate the shape of the $\Pi_{\rm o}$
distribution for much fainter Stokes $I$ flux densities than those
considered in previous work.  The present data have angular resolution
similar to the NVSS, so differences in the measured degree of
polarization because of a difference in resolution are not
expected. Our results can be compared directly with results based on
the NVSS.
 
Noise effects such as those illustrated in Figure~\ref{polstat_demo}
were taken into account by a Monte Carlo analysis.  A set of simulated
catalogs was generated to accurately represent the effects of noise
and the polarized flux density detection threshold in the data.
Stokes $I$ flux densities of simulated sources were drawn from the fit
to observed source counts by \citet{windhorst1990}.  We assume in
these simulations that the Stokes $I$ source counts of polarized
sources have the same shape as those for all radio sources. This is a
reasonable assumption because $\sim 80\%$ of radio sources in the NVSS
display significant polarization \citep{mesaetal2002,tucci2004,bg04}.
The intrinsic Stokes I flux density is multiplied by the degree of
polarization, $\Pi_{\rm o}$, drawn from an assumed $\Pi_{\rm o}$
distribution to obtain the intrinsic polarized intensity $p_{\rm o}$.

The intrinsic Stokes $I_{\rm o}$ and $p_{\rm o}$ of a simulated source
are transformed to observed flux densities $I$ and $p$ by adding noise
with statistical properties identical to the properties of the noise
in the data. First, $p_{\rm o}$ is converted into intrinsic Stokes
$Q_{\rm o}$ and $U_{\rm o}$, assuming a random polarization angle.
The error in the flux density is assumed to consist of a part that is
proportional to the noise in the image at the location of the source,
and a part that is proportional to the flux density of the source,
added in quadrature. The error in the flux density $S$ is evaluated as
\begin{equation}
\sigma_S = S \sqrt { C_1^2 +C_2^2 \Bigl( {\sigma^2 \over S^2} \Bigr)  } \ \ , 
\end{equation}
where $S$ represents the intrinsic flux density $I_{\rm o}$, $Q_{\rm
o}$, or $U_{\rm o}$, $\sigma$ is the rms noise in the image and $C_1$
and $C_2$ are constants. The value of $C_1$ was determined from the
rms variation of the flux density of bright sources in the ten fields
after field registration \citep{taylor2003}.  Flux densities of
sources brighter than $100\,\rm mJy$ varied by $2.5\%$ (rms) over the
ten fields. From this we adopt $C_1 = 0.025$. The value of the
constant $C_2 = 1.3$ was taken to be that found by
\citet{rengelink1997} from Monte Carlo simulations for flux density
errors in the Westerbork Northern Sky Survey (WENSS).  The value of
$\sigma$ is different for each source, to represent variation of the
noise with location in the mosaic.  The distribution of polarized
intensities of the resulting simulated sources, have the same
statistical effects as the observed polarized intensities, including
the effects of noise bias, the detection threshold and the variation
of the noise with position in the mosaic.

\begin{figure}
\begin{center}
\resizebox{14cm}{!}{\includegraphics[angle=0]{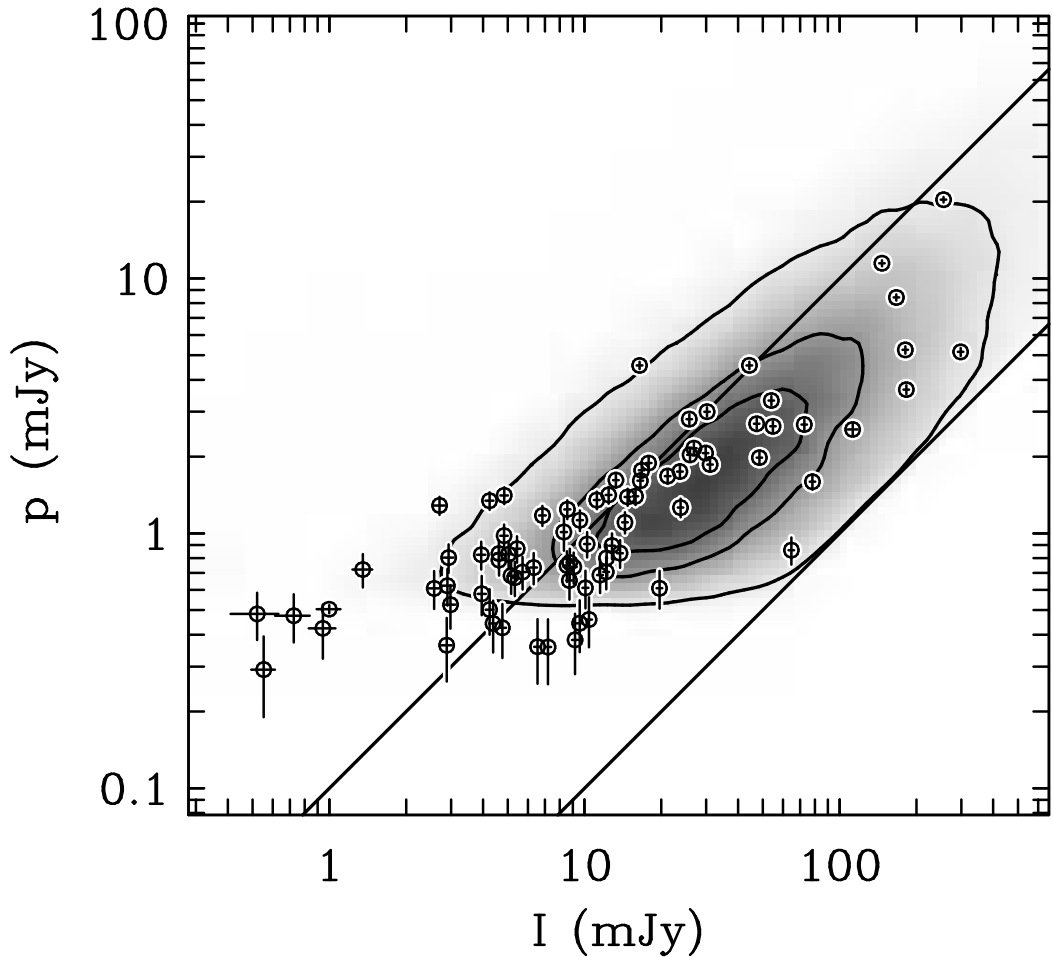}}
\caption{ Distribution in $\log(I)$ - $\log(p)$ of observed sources
(open circles), compared with the simulated distribution that assumes
the $\Pi_{\rm o}$ distribution of \citet{bg04}. Gray scales and contours show
the 2-dimensional probability density function of sources in the
simulated catalog. The inner contours enclose 25\% and 50\% of the simulated
sources, while the outer contour encloses 90\%. Two lines mark
the loci of sources with $\Pi = 1\%$ (right) and $\Pi = 10\%$ (left).
\label{beck-fig}
}  
\end{center}
\end{figure}

To derive the $\Pi_{\rm o}$ distribution directly from the data, we
compare the distribution of the data in a $\log (I) - \log (p)$
diagram with the probability distribution for simulated source
catalogues.  Figure~\ref{beck-fig} shows the observed data points, and
the model probability distribution assuming the $\Pi_{\rm o}$
distribution function of \citet{bg04}.  As expected, sources brighter
than $\sim 80\,\rm mJy$ are represented well by this model. However,
fainter sources in our sample are more highly polarized than predicted
by this distribution.  This is clearly visible in
Figure~\ref{beck-fig} for sources with $10 < I < 30\,\rm mJy$, where
an offset exists between the distribution of observed sources and the
predicted ridge of maximum source density.  A 2-dimensional
Kolmogorov-Smirnov test \citep{peacock1983} rejected the hypothesis
that the data were drawn from the simulated distribution at the 99.9\%
confidence level.

The best fit $\Pi_{\rm o}$ distribution was derived by fitting source
probability density distributions to the data in the $\log(I) -
\log(p)$ plane for trial $\Pi_{\rm o}$ distributions.  The trial
$\Pi_{\rm o}$ distributions were represented by a low-order
Gauss-Hermite series, also called a Gram-Charlier series, following
the description of \citet{vdmarel1993},
\begin{equation}
f(\Pi_{\rm o})=\exp \Bigl({-{\Pi_{\rm o}^2 \over 2\sigma_{\Pi_{\rm o}}^2}}\Bigr) \Bigl[1 + \sum_{i=3}^{N} h_i H_i(\Pi_{\rm o}/\sigma_{\Pi_{\rm o}})  \Bigr].
\label{GHseries-eq}
\end{equation} 
We consider only modest deviations from a Gaussian, since previous
work on bright NVSS sources suggests that the shape of the
distribution is nearly Gaussian.  Assuming that the $\Pi_{\rm o}$
distribution peaks at zero and declines monotonically with increasing
$\Pi_{\rm o}$, we use only the fourth term ($i = 4$) in
Equation~\ref{GHseries-eq}, which results in symmetric deviations.  A
coefficient $h_4> 0$ means that the wings of the distribution are
stronger than the wings of a Gaussian distribution, as shown
graphically by \citet{vdmarel1993}. Higher order terms were not
considered because of the limited size of our data sample at this
time.  The $\Pi_{\rm o}$ distribution defined by
Equation~\ref{GHseries-eq} thus has two free parameters, the Gaussian
standard deviation, $\sigma_{\Pi_{\rm o}}$, and the amplitude of the lowest
order symmetric deviation from a Gaussian, $h_4$.  The fits maximize
the likelihood $L$ of the data as a function of these parameters,
\begin{equation}
L = \prod_{i=1}^{N_{\rm data}} P(I_i,p_i | \sigma_{\Pi_{\rm o}}, h_4), 
\end{equation}
with the probability of an observed ($I_i, p_i$) for a given
$\sigma_{\Pi_{\rm o}}$ and $h_4$
\begin{equation}
P(I_i,p_i | \sigma_{\Pi_{\rm o}}, h_4) = {1 \over M_{\rm model}} \sum_{j=1}^{M_{\rm model}} e^{ {(I_i-I_j)^2 \over 2 \sigma_{I,i}^
2 } + {(p_i-p_j)^2 \over 2 \sigma_{p,i}^2 }    }. 
\end{equation}
The product over $i$ is over all sources in the data, $N_{\rm
data}$. Whereas the sum over $j$ is over all simulated sources in the
catalog, $M_{\rm model}$, for a particular $\sigma_{\Pi_{\rm o}}$ and
$h_4$. Typically $M_{\rm model} \sim 10^5$.  The values $\sigma_{I,i}$
and $\sigma_{p,i}$ are the observed errors in Stokes $I$ and polarized
flux density for the $i^{\rm th}$ source.

\begin{figure}
\begin{center}
\resizebox{14cm}{!}{\includegraphics[angle=0]{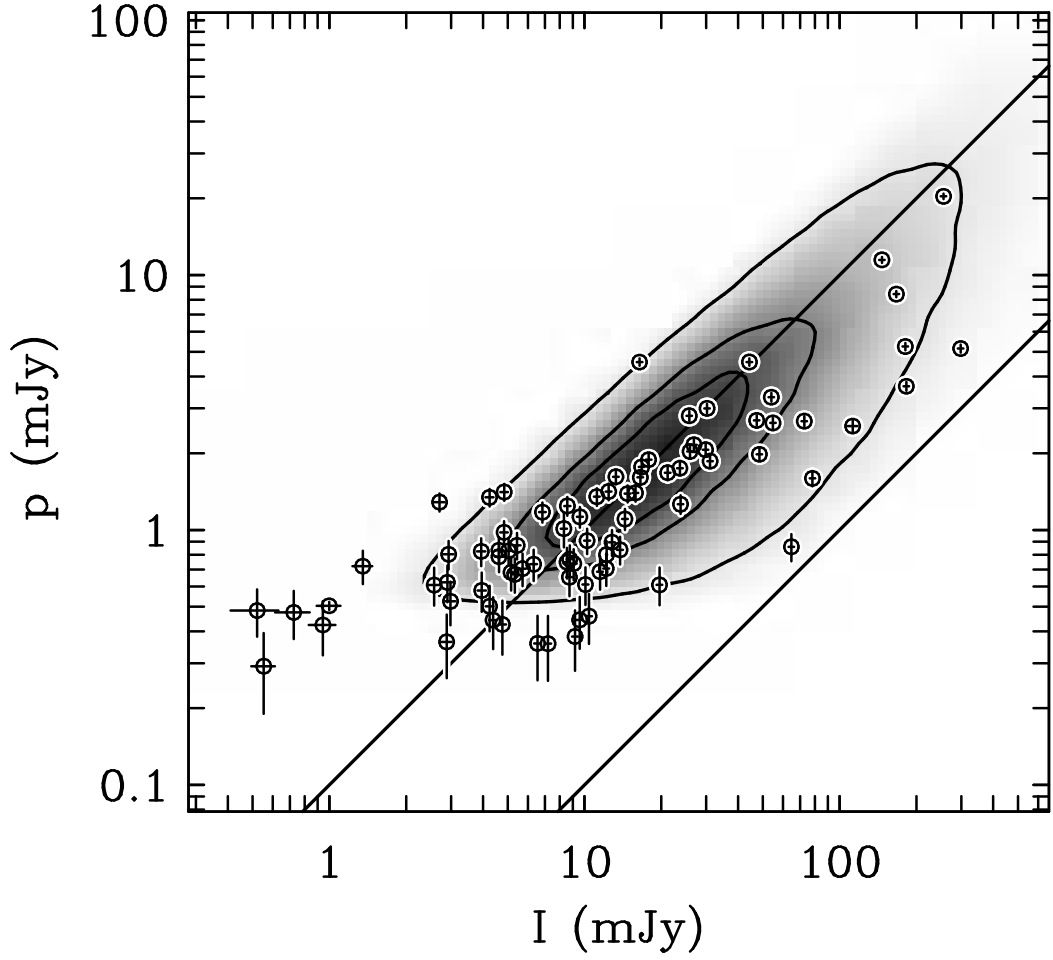}}
\caption{ Same as Figure~\ref{beck-fig}, now for the best fitting Gauss-Hermite
$\Pi_{\rm o}$ distribution, with $\sigma_{\Pi_{\rm o}} = 7.0\%$ and $h_4 = 0.05$.
\label{bestfit_Pi}
}  
\end{center}
\end{figure}

The maximum likelihood $\Pi_{\rm o}$ distribution was found through a
grid search over the parameter space.  The best fitting model has
parameters $\sigma_{\Pi_{\rm o}} = 7\%$, and $h_4 = 0.05$. The
probability density function of this model is shown along with the
data in Figure~\ref{bestfit_Pi}. The uncertainty in the best-fitting
parameters was evaluated empirically. Three hundred randomly selected
samples, each containing on average the same number of sources as the
observed sample, were drawn from the best fitting simulated
catalog. Each of these samples was fitted with the maximum likelihood
fit to evaluate the spread of the best-fitting parameters. Two thirds
of these fits yielded a $\sigma_{\Pi_{\rm o}}$ within 1\% of the
maximum likelihood value $7\%$. We conclude that $\sigma_{\Pi_{\rm o}}
= (7.0 \pm 1.0) \%$. The fitted value of $h_4$ is not independent of
$\sigma_{\Pi_{\rm o}}$, because a larger $h_4$ can partially
compensate for a smaller $\sigma_{\Pi_{\rm o}}$. From the same 300
experimental fits, two thirds yielded a value $h_4 < 0.1$. Although
the data are consistent with $h_4 = 0$, those fits with $h_4$
contrained to be zero yield an average $\sigma_{\Pi_{\rm o}} = (8 \pm
1)\%$. The data therefore suggest that the $\Pi_{\rm o}$ distribution
may be somewhat broader than a simple Gaussian with $\sigma_{\Pi_{\rm
o}} = 7.0\%$.

The best-fitting model was also subjected to a 2-dimensional
Kolmogorov-Smirnov test.  The hypothesis that the data were drawn from
the best-fit distribution in Figure~\ref{bestfit_Pi} was rejected at
the 98\% confidence level. This is much better than the result for the
\citet{bg04} distribution, but it is still suggestive that all the
data are not well represented by the maximum likelihood model
distribution.  This is entirely the result of the fact that the best
fitting Gauss-Hermite distribution does not fit the bright sources in
the sample very well.  We were able to produce a better fit by
creating a set of hybrid simulated catalogs that use the \citet{bg04}
distribution for brighter sources and our best fitting distribution
for faint sources. The transition between these regimes was made
smooth, with equal contributions from the two distributions at a flux
density of 30 mJy.  These hybrid catalogs fitted the data
significantly better, with the best fitting model, using $\sigma_{\Pi_{\rm o}} =
7.0\%$ and $h_4 = 0.05$ at low flux density, passing the 2-dimensional
Kolmogorov-Smirnov test.  The median fractional polarization of the
best fit $\Pi_{\rm o}$ distribution for polarized sources with Stokes
$I$ below $30\,\rm mJy$ is $(4.8 \pm 0.7$)\%.

\subsection{Source Counts}
\label{counts-sec}

To derive source counts of the polarized sources, the effect of the
varying noise level over the mosaic on source detection probability
(completeness correction) was measured as a function of flux density
by repeating the source detection on 1000 simulated images having the
same noise characteristics and source density as the data. Separate
simulations were performed for total intensity and for polarized
intensity images.  Each simulated image contained random Gaussian
noise smoothed to the resolution of the DRAO images, and the same
uniform rms amplitude as the data. For each simulated polarization
image, two independent noise images were created to represent the $Q$
and $U$ images.  Sources were placed at random positions, with flux
densities drawn from the source-count curve derived by
\citet{windhorst1990} between 0.1 and $500\,\rm mJy$.  Sources below
the detection limit were included in the simulations to account for
crowding in the field, and the possibility that faint sources are
observed above the detection limit because of noise.  The resulting
synthetic Stokes $I$ and polarized intensity images were then searched
for sources in an identical manner as for the observed images.

\begin{figure}
\begin{center}
\resizebox{14cm}{!}{\includegraphics[angle=0]{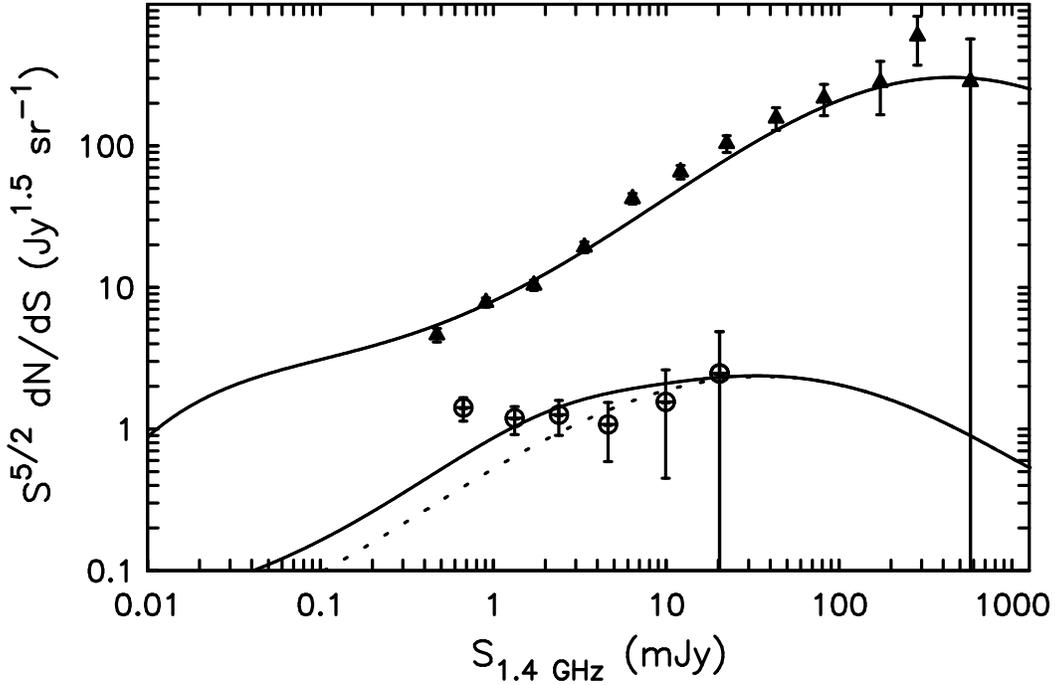}}
\caption{ Euclidean-normalized source counts for total flux density
(triangles) and polarized flux density (circles). The upper solid curve is the
fit to observed source counts from \citet{windhorst1990}.  The lower
solid curve shows polarized source counts predicted by convolving the
Stokes $I$ source counts with the $\Pi$ distribution derived from ELAIS
N1 data for faint sources and the \citet{bg04} distribution for bright
sources, as explained in the text.  The dotted curve shows polarized source
counts derived by convolving only the \citet{bg04} distribution with the
\citet{windhorst1990} source counts curve.
\label{dNdPmodel-fig}
}  
\end{center}
\end{figure}

Figure~\ref{dNdPmodel-fig} shows the derived differential source
counts for total intensity (log $N$ -- log $I$) and polarized flux
density (log $N$ -- log $p$) at $1420\,\rm MHz$, normalized to the
Euclidean expectation in the conventional way. The polarized source
counts are also listed in Table~\ref{polcounts-tab}.  Counts were
derived in bins starting at $500\,\mu$Jy to avoid sources with
completeness correction greater than 10.  This resulted in the removal
of the fifteen faintest sources from the counts.  The Stokes $I$
source counts show good agreement with the Windhorst curve, although
we find somewhat higher numbers around $10\,\rm mJy$. The polarized
source counts are nearly flat in the flux density range observed,
consistent with the increased fractional polarization of the faint
radio sources.

Figure~\ref{dNdPmodel-fig} shows predicted polarized source count
curves derived by convolving Windhorst's Stokes $I$ counts with the
strong-source $\Pi_{\rm o}$ distribution from \citet{bg04} (dashed
curve), as well as our hybrid $\Pi_{\rm o} $ distribution fit to the
$\log I - \log p$ distribution in \S~\ref{maxlike-sec} (solid
curve). The observed polarized source counts show an excess over the
\citet{bg04} extrapolation for $p < 3\,\rm mJy$, and are consistent
with the prediction based on the derived higher $\Pi_{\rm o}$ for
these sources.  The data point at the lowest flux density lies
marginally above our predicted curve.  This may suggest a continuing
trend toward even higher fractional polarization in the sub-mJy
population.  The complete DRAO 30-field survey of the region, and
other deep polarization surveys, will test this.

\subsection{Identification with Spitzer objects}

The positions of the polarized radio sources were examined in the {\sl
Spitzer} SWIRE images of ELAIS N1.  Accurate positions for the sources
were obtained from the VLA FIRST survey images \citep{white1997},
which provides 1.4~GHz continuum Stokes $I$ images at $5\arcsec$
resolution with a 1-$\sigma$ sensitivity of $150\,\mu\rm
Jy~beam^{-1}$.  The sensitivity of the FIRST images is sufficient to
detect a Stokes $I$ counterpart for every polarized source in the
ELAIS N1 deep field.  Approximately 35\% of the polarized radio
sources showed resolved structure on scales of 2\arcsec\ to 30\arcsec\
in the FIRST images.  A polarized source may be associated with a
radio lobe instead of the core of a radio galaxy. Visual inspection
avoided misidentification in such cases. As a comparison to the
polarized sources we also searched the SWIRE images for identification
of sources with Stokes $I$ flux density larger than $1\,\rm mJy$ but
no detectable polarization.

In total, 54 polarized sources were unambiguously identified with {\sl
Spitzer} objects. Another 17 polarized sources had uncertain
identifications, i.e.\ there was more than one possible counterpart
within the errors of the radio position.  Two sources were not covered
by SWIRE.  The remaining 12\% of the sources (10 objects) have no
counterpart in the SWIRE images.  Similar statistics resulted from the
search for counterparts of the Stokes $I$ sources with no detectable
polarized emission.  A wide range in flux density and angular size was
found among the identified {\sl Spitzer} galaxies. Although some faint
galaxies appear unresolved in the {\sl Spitzer} images, the identified
polarized sources seem to be mainly associated with elliptical
galaxies. Three of the polarized sources have counterparts in the {\sl
Spitzer} images that are too faint to appear in the {\sl Spitzer}
ELAIS N1 source catalog.  These sources were not included in the
subsequent analysis.

\begin{figure}
\begin{center}
\resizebox{14cm}{!}{\includegraphics[angle=0]{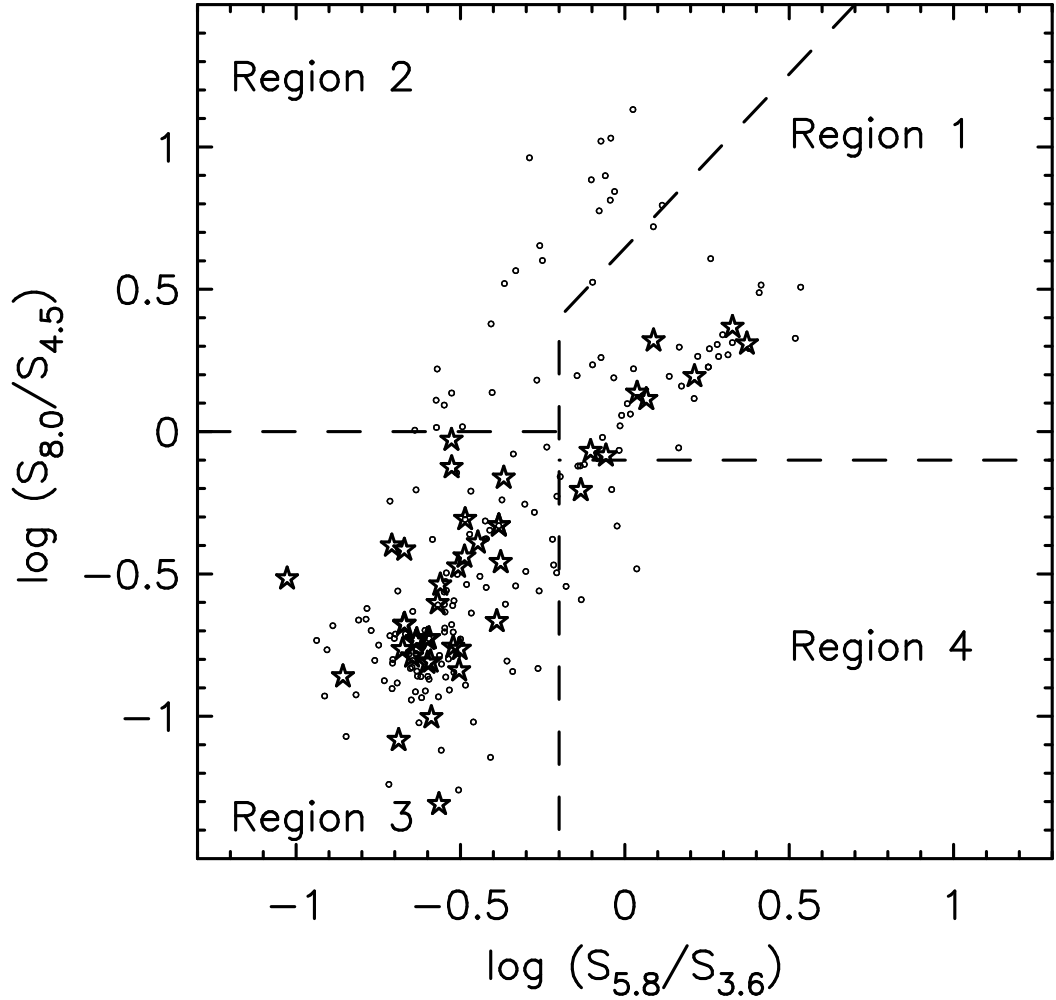}}
\caption{ {\sl Spitzer} near-infrared color-color diagram of ELAIS N1 radio
source host galaxies.  Stars: polarized sources; circles: sources with
no detectable polarized emission. The division of this diagram into four
regions and their interpretation follows \citet{sajina2005} and is
explained in the text.
\label{spitzercol-fig}
}  
\end{center}
\end{figure}

Figure~\ref{spitzercol-fig} shows a near-infrared color-color diagram
of {\sl Spitzer} galaxies identified with radio sources that had
catalogued flux densities in all four IRAC bands at 3.6, 4.5, 5.8, and
$8.0\,\mu$m.  This includes 41 of the polarized sources.  Since many
of the sources are identified with extended galaxies, we used the
isophotal fluxes in each band.  Our analysis of this color-color
diagram is based on the modeling of \citet{sajina2005}, which divided
the diagram into the four regions separated by dashed lines in
Figure~\ref{spitzercol-fig}.  Boundaries between these regions were
defined so as to separate galaxies depending on the strength of
near-infrared PAH bands and the slope of the near-infrared
continuum. Region 1 is mainly populated by sources with a continuum
that rises with wavelength. The near-infrared spectrum of these
sources is usually dominated by non-equilibrium emission of
stochastically heated very small dust grains, interpreted as PAH
destruction by the hard ultraviolet spectrum of an AGN.  Region 2 is
occupied mainly by dusty star-forming galaxies with strong PAH bands
at red-shift $z < 0.5$. This is because the strongest PAH features at
low redshift contribute to the flux in IRAC bands at 3.6 and
$8.0\,\mu$m. Region 3 is occupied by galaxies with fainter PAH
emission, or by dusty starforming galaxies at redshift $z = 0.5 -
1.5$. A conspicuous concentration of galaxies in region 3 is the blue
clump near $\log(S_{5.8})/\log(S_{3.6}) = -0.6$,
$\log(S_{8.0})/\log(S_{4.5}) = -0.8$.  Galaxies in the blue clump have
a near-infrared continuum that declines with wavelength, since the
spectrum is dominated by the starlight of an old stellar
population. These are elliptical galaxies, at a wide range of
redshift.  Region 4 is populated by PAH-dominated sources at redshift
$z = 1.5 - 2$.

The number of sources by region in Figure~\ref{spitzercol-fig} is
listed in Table~\ref{table_irac_colcol}. The host galaxies of the
polarized sources occupy mainly regions 1 and 3. The source in
region 4 cannot be considered a convincing high-redshift galaxy in
view of uncertainties in the photometry and its proximity to the
boundary with regions 1 and 3. The majority (71\%) of the host
galaxies of polarized sources in the {\sl Spitzer} color-color diagram
are either in region 1 or in the blue clump. In both cases, the radio
emission is interpreted as emission from an AGN.

Eleven polarized sources (27\%) are found in the region of
PAH-dominated galaxies, with $\log(S_{8.0})/\log(S_{4.5}) > -0.5$, a
few dex above the blue clump. Their location in the
color-color diagram suggests either PAH-dominated galaxies at redshift
$0.5 - 1.5$, or galaxies with faint PAH bands at lower redshift
\citep{sajina2005}. Inspection of the FIRST and SWIRE images of these
sources shows that all appear to be elliptical galaxies with a smooth
morphology, and some have resolved symmetric radio sources, suggestive
of radio lobes. This suggests that the polarized radio emission in
these galaxies is associated with AGN activity, while the PAH
emission originates from dust at a substantial distance from the AGN,
where it is shielded from the hard ultraviolet radiation that would
destroy the PAHs. Dust in elliptical galaxies is a common phenomenon
\citep{goudfrooij1994}. It can be produced in the envelopes of cool
red giant stars or it can be acquired through a merger with a gas-rich
galaxy.

The Stokes $I$ sources with no detectable polarized emission generally
occupy the same areas of the color-color diagram as the polarized
sources, with two exceptions. First, a significant fraction (15\%) of
the Stokes $I$ sources is located in region 2, the area where star
forming, PAH-dominated galaxies are expected. These objects are
generally the faintest Stokes I radio sources, and none would be
detectable in polarization at our sensitivity level.  The Stokes $I$
sources in region 2 are likely members of a population of star-forming
galaxies that is believed to make up a large fraction of the radio
source population below $\sim$1 mJy.  The second exception is that
Table~\ref{table_irac_colcol} indicates an excess of polarized sources
(27\%) relative to sources with no detected polarization (14\%) in
region 3b, associated with galaxies having PAH emission. This
difference is significant if Poisson errors are assumed. Confirmation
will require a more complete sample of polarized radio sources
identified with {\sl Spitzer} galaxies.

\subsection{Nature of the mJy polarized source population}

The polarized sources found in the ELAIS N1 field have a median Stokes
$I$ flux density of 12 mJy. Models of radio source populations fitted
to the total radio source counts suggest that most radio sources with
$1420\,\rm MHz$ flux density $\gtrsim 1\,\rm mJy$ are steep-spectrum
radio galaxies, whose power is ultimately derived from accretion onto
a compact object.  However, it is not clear a priori that a faint
polarization-selected sample of radio sources is representative of the
entire population.  In principle, a highly polarized population of
faint radio sources may constitute a significant fraction of a sample
of faint polarized sources.

In Section~\ref{maxlike-sec} we presented evidence that faint
extragalactic polarized radio sources are on average more highly
polarized than bright sources, with a median fractional polarization
approximately three times higher.  From an analysis of sources with
Stokes $I > 100$ mJy in the NVSS, \citet{mesaetal2002} also noted that
the median fractional polarization of radio sources increases with
decreasing flux density, from 1.05\% for Stokes $ I > 800$ mJy to
1.84\% between $100-200\,\rm mJy$.  From a similar analysis of the
NVSS, \citet{tucci2004} showed that the anti-correlation between flux
density and percentage polarization occurs only for steep-spectrum
sources. They found that the median percentage polarization for
steep-spectrum sources increased from 1.14\% for flux densities
greater than $800\,\rm mJy$ to 1.77\% between $100 - 200\,\rm mJy$.
Our result extends this to much lower flux densities and indicates a
much stronger effect for faint sources, resulting in a median
polarization of $4.8\%$ at Stokes $I = 10 - 30\,\rm mJy$.  Analysis
of Table~\ref{table_polsources} shows that these faint polarized
emitters are dominated by steep-spectrum sources; all but one of the
polarized sources for which a 325 --$1420\,\rm MHz$ spectral index
exists (75\%) has $\alpha < -0.4$.
 
Polarized sources that can be identified with galaxies in the {\sl
Spitzer} ELAIS N1 deep field, are associated with elliptical
galaxies. Most of the host galaxies have near-infrared colors typical
for dust emission from the vicinity of an AGN, or an old stellar
population with no significant dust emission. The majority of the
polarized sources is associated with AGN activity for this reason.
The remaining 11 polarized sources associated with galaxies having PAH
emission are also likely to contain an AGN.  Some are clearly resolved
double-lobed objects in the FIRST images, and all appear to be
elliptical galaxies with a smooth brightness distribution.  We
conclude that there is no evidence for galaxies in our sample of
polarized sources in which the radio emission is powered by star
formation.  The higher fractional polarization of faint radio sources
may be related to a population of radio-quiet AGN in which fainter
radio emission correlates with conditions that favor increased
polarization, for example more ordered magnetic fields or less
internal Faraday depolarization.  High-resolution polarimetry of these
objects will provide more insight into their nature.

\section{Conclusions}

We present sensitive observations of a complete sample of compact
polarized radio sources, as part of a deep integration of the ELAIS N1
region made with the Synthesis Telescope at the Dominion Radio
Astrophysical Observatory. A total of 83 polarized sources was
detected in the ten-field mosaic.

The distribution of fractional polarization of faint polarized sources
was investigated with a Monte-Carlo analysis that generates synthetic
source lists with the same noise statistics and observational
selection criteria as the data. Maximum-likelihood fits of the
synthetic source lists to the data in the $\log(I)$ - $\log(p)$ plane
yielded a best fitting Gauss-Hermite function with $\sigma_{\Pi_{\rm o}} = (7.0\
\pm\ 1.0)\%$, $h_4 = 0.05\ \pm 0.05$ for the distribution of intrinsic
fractional polarization. The data demonstrate a trend of increasing
fractional polarization with decreasing flux density.

Polarized source counts from the ELAIS N1 deep field are presented
down to $0.5\,\rm mJy$. We find that the Euclidean-normalized
polarized counts remain flat below $1\,\rm mJy$. The distribution of
fractional polarization derived from our Monte Carlo analysis is
convolved with the total-intensity source counts to produce a
prediction of the polarized source counts.  The predicted
Euclidean-normalized polarized counts are nearly flat to $\sim 2\,\rm
mJy$, in good agreement with the data.  However, the data at the
faintest polarized flux densities suggests a continuing trend of
increased polarization fraction with decreasing flux density.

The near-infrared color-color diagram for host galaxies identified
with the polarized sources in the ELAIS N1 field shows that most of
the host galaxies are ellipticals, or galaxies for which the
near-infrared spectrum is dominated by stochastically heated very
small grains, presumably from the vicinity of an AGN. Some of the host
galaxies appear to have PAH bands in their near-infrared spectrum, but
the morphological resemblance with ellipticals, and the fact that some
of these polarized sources are resolved radio galaxies in the FIRST
survey, indicates that these objects also harbour AGN. We suggest that
the higher degree of polarization indicates a difference between AGN
observed at a flux density of hundreds of mJy, and fainter AGN.

\section*{Acknowledgments}

Observations and research on the DRAO Planck Deep Fields are supported
by the Natural Sciences and Engineering Research Council of Canada and
the National Research Council Canada. Ev Sheehan of DRAO was
instrumental in improving the sensitivity of the DRAO Synthesis
Telescope, with the outcome recorded in this paper. We are indebted to
him for his skill and dedication to this difficult task.  The Dominion
Radio Astrophysical Observatory is operated as a National Facility by
the National Research Council of Canada.

{}
\clearpage

\begin{figure}[h]
\begin{center}
\caption{ {\bf [This Figure is provided as a separate image f2.gif]}
Continuum images of the ELAIS N1 field at 21 cm wavelength. Top row:
Stokes I (gray scales linear from $-0.1$ to $+5$ mJy
beam$^{-1}$). Middle row: Stokes $Q$ (gray scales linear from $-1$ to
$+1$ mJy beam$^{-1}$). Bottom row: Stokes $U$ (gray scales linear from
$-1$ to $+1$ mJy beam$^{-1}$). Panels on the right show an enlargement
of the area indicated by the white frame in the Stokes $Q$ image.
\label{EN21-fig}
}  
\end{center}
\end{figure}

\begin{figure}
\begin{center}
\caption{ {\bf [This Figure is provided as a separate image f3.gif]}
The distribution of amplitudes in the $Q$ and $U$ images after
dividing by the mosaic weights to produce an image with uniform noise
over the map equal to the noise value at the map center.  Gaussian
fits to the distributions (solid curves) were used to measure the map
center rms at $78\,\mu$Jy.
\label{fig:Q_Unoise}
}  
\end{center}
\end{figure}

\begin{deluxetable}{cccrrrrrr}
\tabletypesize{\scriptsize}
\tablecaption{ ELAIS N1 Polarized Sources \label{table_polsources} }
\tablehead{
\colhead{Source} & \colhead{RA} & \colhead{DEC} & \colhead{$I$} & \colhead{$p_{o}$} & 
\colhead{PA} & \colhead{$\Pi_0$}  & \multicolumn{2}{c}{$\alpha_{325-1420}$} \\
\colhead{Number} & \multicolumn{2}{c}{(J2000)}  &\colhead{(mJy)}&\colhead{(mJy beam$^{-1}$)} &  
\colhead{($^{\circ}$)}  & \colhead{(\%)} & & \\
}
\startdata
 1 & 16 02 05.52 $\pm$  0.29 & 54 54 43.2 $\pm$   3.0 &     16.82 $\pm$  0.56 &    1.74 $\pm$  0.11 &      26 &  10.4  $\pm$   0.7 &  $-$0.78 $\pm$  0.06 \\ 
 2 & 16 02 34.37 $\pm$  0.48 & 54 54 01.8 $\pm$   4.2 &      7.95 $\pm$  0.40 &    0.98 $\pm$  0.16 &      37 &  11.8  $\pm$   1.9 &  $-$0.75 $\pm$  0.11 \\ 
 3 & 16 02 42.00 $\pm$  0.44 & 55 10 01.2 $\pm$   4.4 &      8.51 $\pm$  0.78 &    0.78 $\pm$  0.10 &      71 &  19.9  $\pm$   2.7 &  $-$0.23 $\pm$  0.22 \\ 
 4 & 16 03 11.52 $\pm$  0.34 & 55 39 07.2 $\pm$   3.8 &      4.65 $\pm$  0.40 &    1.37 $\pm$  0.11 &      27 &   28.2 $\pm$   2.4 & $>-$0.64 \hfill \quad & \\ 
 5 & 16 03 34.08 $\pm$  0.44 & 54 29 20.4 $\pm$   3.9 &      4.53 $\pm$  0.37 &    0.79 $\pm$  0.10 &      36 &  17.2  $\pm$   2.3 &  $-$1.46 $\pm$  0.07 \\ 
 6 & 16 04 42.24 $\pm$  0.14 & 54 38 45.6 $\pm$   1.4 &     19.42 $\pm$  0.55 &    1.88 $\pm$  0.11 &      33 &  10.5  $\pm$   0.7 &  $-$0.99 $\pm$  0.04 \\ 
 7 & 16 05 05.52 $\pm$  0.15 & 55 00 46.8 $\pm$   1.4 &     27.86 $\pm$  0.74 &    1.25 $\pm$  0.11 &       7 &   5.3  $\pm$   0.5 &  $-$1.33 $\pm$  0.03 \\ 
 8 & 16 05 23.04 $\pm$  0.20 & 54 29 31.2 $\pm$   1.8 &      6.39 $\pm$  0.29 &    1.16 $\pm$  0.11 &   $-$12 &   17.0 $\pm$   1.6 & $>-$0.43 \hfill \quad & \\ 
 9 & 16 05 38.88 $\pm$  0.04 & 54 39 28.8 $\pm$   0.4 &    187.61 $\pm$  4.74 &    8.42 $\pm$  0.24 &      21 &   5.0  $\pm$   0.2 &  $-$1.00 $\pm$  0.03 \\ 
10 & 16 05 38.88 $\pm$  0.20 & 54 41 27.6 $\pm$   1.9 &      2.00 $\pm$  0.35 &    1.28 $\pm$  0.11 &   $-$37 &   47.3 $\pm$   4.5 & $>-$1.22 \hfill \quad & \\ 
11 & 16 06 01.44 $\pm$  0.04 & 54 54 10.8 $\pm$   0.5 &    209.87 $\pm$  5.29 &    5.24 $\pm$  0.17 &      15 &   2.9  $\pm$   0.1 &  $-$0.91 $\pm$  0.03 \\ 
12 & 16 06 08.40 $\pm$  0.15 & 55 16 04.8 $\pm$   1.6 &      5.06 $\pm$  0.28 &    0.86 $\pm$  0.10 &      16 &  15.8  $\pm$   2.0 &  $-$0.69 $\pm$  0.19 \\ 
13 & 16 06 13.44 $\pm$  0.23 & 55 01 58.8 $\pm$   2.3 &      1.14 $\pm$  0.18 &    0.71 $\pm$  0.11 &   $-$14 &   52.6 $\pm$   9.3 & $>-$1.59 \hfill \quad & \\ 
14 & 16 06 35.52 $\pm$  0.10 & 54 35 02.4 $\pm$   1.0 &     15.96 $\pm$  0.51 &    1.61 $\pm$  0.11 &      44 &  12.1  $\pm$   0.9 &  $-$0.73 $\pm$  0.06 \\ 
15 & 16 06 58.80 $\pm$  0.32 & 54 43 12.0 $\pm$   3.0 &      4.31 $\pm$  0.21 &    0.41 $\pm$  0.10 &    $-$9 &    8.6 $\pm$   2.2 & $>-$0.69 \hfill \quad & \\ 
16 & 16 07 22.56 $\pm$  0.22 & 55 31 04.8 $\pm$   2.8 &     12.16 $\pm$  0.45 &    0.73 $\pm$  0.10 &       1 &   8.0  $\pm$   1.2 &  $-$1.05 $\pm$  0.05 \\ 
17 & 16 08 21.36 $\pm$  0.11 & 56 13 51.6 $\pm$   0.8 &    221.05 $\pm$  5.59 &    3.65 $\pm$  0.15 &      54 &   2.0  $\pm$   0.1 &   0.30 $\pm$  0.03 \\ 
18 & 16 08 28.56 $\pm$  0.28 & 54 10 37.2 $\pm$   3.7 &     19.94 $\pm$  0.56 &    0.59 $\pm$  0.10 &      26 &   3.0  $\pm$   0.5 &  $-$0.68 $\pm$  0.06 \\ 
19 & 16 08 38.64 $\pm$  0.29 & 54 14 34.8 $\pm$   3.3 &      2.41 $\pm$  0.21 &    0.51 $\pm$  0.10 &   $-$19 &   17.0 $\pm$   3.5 & $>-$1.09 \hfill \quad & \\ 
20 & 16 08 47.76 $\pm$  0.20 & 56 11 16.8 $\pm$   1.6 &     27.09 $\pm$  1.03 &    1.66 $\pm$  0.11 &       8 &   7.8  $\pm$   0.6 &  $-$1.19 $\pm$  0.03 \\ 
21 & 16 08 58.32 $\pm$  0.36 & 55 56 27.6 $\pm$   2.2 &      9.29 $\pm$  0.48 &    0.69 $\pm$  0.10 &      69 &  12.0  $\pm$   1.8 &  $-$0.82 $\pm$  0.09 \\ 
22 & 16 09 04.32 $\pm$  0.26 & 56 10 33.6 $\pm$   2.2 &     12.68 $\pm$  0.60 &    1.33 $\pm$  0.11 &      23 &  11.9  $\pm$   1.0 &  $-$1.13 $\pm$  0.05 \\ 
23 & 16 09 11.04 $\pm$  0.18 & 55 26 31.2 $\pm$   2.2 &      4.51 $\pm$  0.22 &    0.78 $\pm$  0.10 &       4 &  16.9  $\pm$   2.3 &  $-$0.86 $\pm$  0.17 \\ 
24 & 16 09 22.80 $\pm$  0.15 & 56 15 03.6 $\pm$   1.4 &     32.46 $\pm$  0.95 &    2.14 $\pm$  0.12 &      21 &   8.0  $\pm$   0.5 &  $-$0.63 $\pm$  0.04 \\ 
25 & 16 09 31.68 $\pm$  0.12 & 55 25 04.8 $\pm$   1.3 &     15.32 $\pm$  0.47 &    1.41 $\pm$  0.11 &      25 &  11.4  $\pm$   0.9 &  $-$1.05 $\pm$  0.05 \\ 
26 & 16 09 36.24 $\pm$  0.18 & 55 27 03.6 $\pm$   1.5 &     10.95 $\pm$  0.37 &    1.12 $\pm$  0.10 &       8 &  11.7  $\pm$   1.1 &  $-$0.71 $\pm$  0.09 \\ 
27 & 16 09 44.40 $\pm$  0.38 & 54 37 51.6 $\pm$   2.9 &      6.61 $\pm$  0.24 &    0.35 $\pm$  0.10 &      59 &    5.3 $\pm$   1.6 & $>-$0.40 \hfill \quad & \\ 
28 & 16 09 52.56 $\pm$  0.24 & 55 07 08.4 $\pm$   2.5 &      0.68 $\pm$  0.14 &    0.42 $\pm$  0.10 &       1 &   44.1 $\pm$  12.0 & $>-$1.95 \hfill \quad & \\ 
29 & 16 10 03.12 $\pm$  0.10 & 55 52 37.2 $\pm$   1.1 &     96.83 $\pm$  2.47 &    1.59 $\pm$  0.11 &      30 &   2.0  $\pm$   0.2 &  $-$0.79 $\pm$  0.03 \\ 
30 & 16 10 27.12 $\pm$  0.14 & 54 12 54.0 $\pm$   1.6 &      8.51 $\pm$  0.28 &    1.24 $\pm$  0.11 &      30 &  14.4  $\pm$   1.3 &  $-$0.43 $\pm$  0.17 \\ 
31 & 16 10 57.84 $\pm$  0.08 & 55 35 24.0 $\pm$   0.7 &     19.46 $\pm$  0.53 &    1.61 $\pm$  0.11 &       5 &   9.7  $\pm$   0.7 &  $-$0.63 $\pm$  0.06 \\ 
32 & 16 11 00.48 $\pm$  0.06 & 54 42 03.6 $\pm$   0.6 &     29.41 $\pm$  0.76 &    2.03 $\pm$  0.11 &      19 &   7.8  $\pm$   0.5 &  $-$0.74 $\pm$  0.04 \\ 
33 & 16 11 20.40 $\pm$  0.08 & 55 28 44.4 $\pm$   0.9 &     18.22 $\pm$  0.50 &    1.39 $\pm$  0.11 &       3 &   8.8  $\pm$   0.7 &  $-$0.88 $\pm$  0.05 \\ 
34 & 16 11 21.12 $\pm$  0.30 & 54 31 55.2 $\pm$   3.3 &      3.62 $\pm$  0.34 &    0.43 $\pm$  0.10 &    $-$5 &    9.9 $\pm$   2.4 & $>-$0.81 \hfill \quad & \\ 
35 & 16 11 29.04 $\pm$  0.29 & 55 51 36.0 $\pm$   2.6 &     13.22 $\pm$  0.41 &    0.44 $\pm$  0.10 &      25 &   4.3  $\pm$   1.0 &  $-$0.87 $\pm$  0.06 \\ 
36 & 16 11 37.92 $\pm$  0.29 & 53 59 34.8 $\pm$   4.2 &     13.82 $\pm$  0.44 &    0.82 $\pm$  0.10 &   $-$18 &   5.9  $\pm$   0.8 &  $-$1.03 $\pm$  0.05 \\ 
37 & 16 11 38.16 $\pm$  0.26 & 55 59 52.8 $\pm$   2.7 &      4.50 $\pm$  0.23 &    0.48 $\pm$  0.10 &      14 &  11.3  $\pm$   2.4 &  $-$1.19 $\pm$  0.11 \\ 
38 & 16 11 50.88 $\pm$  0.22 & 55 00 54.0 $\pm$   1.0 &      9.14 $\pm$  0.41 &    0.82 $\pm$  0.10 &      14 &  16.3  $\pm$   2.1 &  $-$0.63 $\pm$  0.12 \\ 
39 & 16 12 12.24 $\pm$  0.02 & 55 22 48.0 $\pm$   0.2 &    312.36 $\pm$  7.86 &   20.37 $\pm$  0.52 &      29 &   8.0  $\pm$   0.3 &  $-$1.11 $\pm$  0.03 \\ 
40 & 16 12 24.00 $\pm$  0.13 & 55 26 02.4 $\pm$   1.5 &      9.57 $\pm$  0.34 &    0.74 $\pm$  0.10 &      32 &   8.7  $\pm$   1.2 &  $-$0.56 $\pm$  0.12 \\ 
41 & 16 12 28.56 $\pm$  0.34 & 55 06 46.8 $\pm$   2.0 &     10.06 $\pm$  0.30 &    0.37 $\pm$  0.10 &      59 &   4.1  $\pm$   1.1 &  $-$1.13 $\pm$  0.06 \\ 
42 & 16 12 31.68 $\pm$  0.31 & 54 18 10.8 $\pm$   2.5 &      9.81 $\pm$  0.36 &    0.43 $\pm$  0.10 &      21 &   4.5  $\pm$   1.1 &  $-$0.73 $\pm$  0.10 \\ 
43 & 16 12 35.28 $\pm$  0.04 & 56 28 19.2 $\pm$   0.4 &    176.88 $\pm$  4.51 &   11.47 $\pm$  0.31 &      39 &   7.8  $\pm$   0.3 &  $-$0.96 $\pm$  0.03 \\ 
44 & 16 12 47.52 $\pm$  0.33 & 55 02 31.2 $\pm$   2.9 &      7.68 $\pm$  0.25 &    0.35 $\pm$  0.10 &      51 &   4.9  $\pm$   1.4 &  $-$0.74 $\pm$  0.12 \\ 
45 & 16 12 51.36 $\pm$  0.20 & 56 03 50.4 $\pm$   2.4 &      2.59 $\pm$  0.19 &    0.61 $\pm$  0.10 &      25 &   20.9 $\pm$   3.7 & $>-$1.04 \hfill \quad & \\ 
46 & 16 13 02.40 $\pm$  0.15 & 54 32 27.6 $\pm$   1.3 &      6.48 $\pm$  0.25 &    0.73 $\pm$  0.10 &      56 &  11.5  $\pm$   1.7 &  $-$0.86 $\pm$  0.12 \\ 
47 & 16 13 16.80 $\pm$  0.36 & 56 08 13.2 $\pm$   3.8 &      3.85 $\pm$  0.36 &    0.56 $\pm$  0.10 &   $-$28 &   14.1 $\pm$   2.6 & $>-$0.77 \hfill \quad & \\ 
48 & 16 13 19.20 $\pm$  0.09 & 54 16 40.8 $\pm$   1.0 &      5.78 $\pm$  0.43 &    1.34 $\pm$  0.11 &       5 &   31.5 $\pm$   2.8 & $>-$0.50 \hfill \quad & \\ 
49 & 16 13 25.92 $\pm$  0.24 & 55 39 39.6 $\pm$   2.4 &      3.06 $\pm$  0.18 &    0.35 $\pm$  0.10 &      17 &   12.2 $\pm$   3.6 & $>-$0.93 \hfill \quad & \\ 
50 & 16 13 26.64 $\pm$  0.10 & 55 15 46.8 $\pm$   1.1 &     14.60 $\pm$  0.14 &    0.80 $\pm$  0.10 &      14 &   6.5  $\pm$   0.9 &  $-$1.19 $\pm$  0.04 \\ 
51 & 16 13 28.80 $\pm$  0.27 & 56 17 49.2 $\pm$   2.1 &     17.29 $\pm$  0.57 &    0.87 $\pm$  0.10 &      26 &   6.8  $\pm$   0.8 &  $-$0.53 $\pm$  0.07 \\ 
52 & 16 13 30.72 $\pm$  0.06 & 54 27 21.6 $\pm$   0.6 &     80.29 $\pm$  2.03 &    2.67 $\pm$  0.12 &      15 &   3.7  $\pm$   0.2 &  $-$0.85 $\pm$  0.03 \\ 
53 & 16 13 36.72 $\pm$  0.36 & 54 11 16.8 $\pm$   2.7 &      1.70 $\pm$  0.36 &    0.59 $\pm$  0.10 &       1 &   22.9 $\pm$   4.2 & $>-$1.33 \hfill \quad & \\ 
54 & 16 13 41.76 $\pm$  0.19 & 56 11 49.2 $\pm$   1.4 &    101.05 $\pm$  7.62 &    2.80 $\pm$  0.12 &      22 &  10.9  $\pm$   0.6 &   0.09 $\pm$  0.04 \\ 
55 & 16 13 48.48 $\pm$  0.27 & 54 14 13.2 $\pm$   2.5 &      8.12 $\pm$  0.40 &    0.64 $\pm$  0.10 &      18 &   7.3  $\pm$   1.2 &  $-$1.10 $\pm$  0.07 \\ 
56 & 16 13 56.16 $\pm$  0.25 & 54 57 28.8 $\pm$   2.5 &      0.45 $\pm$  0.09 &    0.28 $\pm$  0.10 &    $-$1 &  50.6  $\pm$  19.2 &  $-$3.97 $\pm$  0.03 \\ 
57 & 16 13 56.84 $\pm$  0.22 & 55 02 08.2 $\pm$   2.3 &      0.81 $\pm$  0.13 &    0.50 $\pm$  0.02 &   $-$31 &   49.8 $\pm$   5.7 & $>-$1.82 \hfill \quad & \\ 
58 & 16 14 00.96 $\pm$  0.20 & 53 57 21.6 $\pm$   1.8 &     14.75 $\pm$  0.49 &    1.08 $\pm$  0.10 &      32 &   7.5  $\pm$   0.8 &  $-$0.41 $\pm$  0.10 \\ 
59 & 16 14 16.80 $\pm$  0.29 & 55 42 57.6 $\pm$   3.5 &      1.14 $\pm$  0.36 &    0.46 $\pm$  0.10 &      56 &   64.0 $\pm$  17.2 & $>-$1.60 \hfill \quad & \\ 
60 & 16 14 21.12 $\pm$  0.06 & 55 36 39.6 $\pm$   0.6 &     35.41 $\pm$  0.93 &    3.00 $\pm$  0.13 &      18 &   9.9  $\pm$   0.5 &  $-$0.74 $\pm$  0.04 \\ 
61 & 16 14 32.64 $\pm$  0.20 & 55 38 31.2 $\pm$   2.5 &      1.27 $\pm$  0.39 &    0.47 $\pm$  0.10 &   $-$79 &   90.8 $\pm$  27.5 & $>-$1.52 \hfill \quad & \\ 
62 & 16 15 27.36 $\pm$  0.13 & 54 27 10.8 $\pm$   1.2 &      4.51 $\pm$  0.20 &    0.97 $\pm$  0.10 &       5 &   20.1 $\pm$   2.3 & $>-$0.66 \hfill \quad & \\ 
63 & 16 15 30.96 $\pm$  0.16 & 54 52 30.0 $\pm$   2.0 &      5.43 $\pm$  0.22 &    0.68 $\pm$  0.10 &      20 &  13.1  $\pm$   2.0 &  $-$1.16 $\pm$  0.09 \\ 
64 & 16 15 36.72 $\pm$  0.19 & 53 46 37.2 $\pm$   2.4 &     54.20 $\pm$  1.46 &    2.64 $\pm$  0.12 &      19 &   5.6  $\pm$   0.3 &  $-$0.85 $\pm$  0.03 \\ 
65 & 16 15 49.68 $\pm$  0.06 & 55 16 40.8 $\pm$   0.7 &     27.80 $\pm$  0.72 &    1.75 $\pm$  0.11 &      35 &   7.4  $\pm$   0.5 &  $-$0.68 $\pm$  0.04 \\ 
66 & 16 16 23.04 $\pm$  0.14 & 55 27 00.0 $\pm$   1.3 &     13.10 $\pm$  0.40 &    0.90 $\pm$  0.10 &      27 &   8.8  $\pm$   1.0 &  $-$0.89 $\pm$  0.06 \\ 
67 & 16 16 23.52 $\pm$  0.17 & 54 57 43.2 $\pm$   1.6 &     10.35 $\pm$  0.48 &    0.60 $\pm$  0.10 &      10 &   6.0  $\pm$   1.0 &  $-$0.97 $\pm$  0.07 \\ 
68 & 16 16 37.92 $\pm$  0.18 & 55 45 14.4 $\pm$   2.1 &     74.66 $\pm$  1.90 &    0.85 $\pm$  0.10 &      51 &   1.3  $\pm$   0.2 &  $-$0.61 $\pm$  0.03 \\ 
69 & 16 16 39.36 $\pm$  0.11 & 53 58 12.0 $\pm$   0.9 &    351.23 $\pm$  9.13 &    5.14 $\pm$  0.19 &      23 &   1.7  $\pm$   0.1 &  $-$0.85 $\pm$  0.03 \\ 
70 & 16 16 40.08 $\pm$  0.18 & 56 20 38.4 $\pm$   1.5 &     18.64 $\pm$  0.70 &    1.37 $\pm$  0.11 &      24 &   9.3  $\pm$   0.8 &  $-$0.52 $\pm$  0.07 \\ 
71 & 16 17 57.60 $\pm$  0.22 & 54 51 36.0 $\pm$   3.0 &     14.65 $\pm$  0.46 &    0.68 $\pm$  0.10 &   $-$17 &   5.9  $\pm$   0.9 &  $-$1.11 $\pm$  0.04 \\ 
72 & 16 18 06.72 $\pm$  0.36 & 54 42 46.8 $\pm$   2.4 &      5.31 $\pm$  0.26 &    0.65 $\pm$  0.10 &      24 &   12.3 $\pm$   2.0 & $>-$0.55 \hfill \quad & \\ 
73 & 16 18 32.64 $\pm$  0.05 & 54 31 44.4 $\pm$   0.5 &     48.68 $\pm$  1.27 &    4.56 $\pm$  0.15 &    $-$1 &  10.3  $\pm$   0.4 &  $-$0.78 $\pm$  0.03 \\ 
74 & 16 18 57.57 $\pm$  0.12 & 54 29 26.2 $\pm$   1.3 &    132.06 $\pm$  5.88 &    2.55 $\pm$  0.14 &      21 &   2.3  $\pm$   0.2 &  $-$1.39 $\pm$  0.03 \\ 
75 & 16 18 59.28 $\pm$  0.18 & 54 52 40.8 $\pm$   1.7 &     40.08 $\pm$  1.07 &    1.86 $\pm$  0.11 &      21 &   6.0  $\pm$   0.4 &  $-$0.50 $\pm$  0.04 \\ 
76 & 16 19 15.36 $\pm$  0.20 & 55 05 13.2 $\pm$   1.4 &     16.65 $\pm$  0.50 &    0.69 $\pm$  0.10 &    $-$2 &   5.7  $\pm$   0.9 &  $-$0.30 $\pm$  0.11 \\ 
77 & 16 19 19.20 $\pm$  0.08 & 55 35 56.4 $\pm$   0.8 &     53.48 $\pm$  1.36 &    1.98 $\pm$  0.11 &      15 &   4.1  $\pm$   0.3 &  $-$0.44 $\pm$  0.04 \\ 
78 & 16 19 19.44 $\pm$  0.06 & 54 48 25.2 $\pm$   0.6 &     57.59 $\pm$  1.46 &    3.32 $\pm$  0.13 &      11 &   6.1  $\pm$   0.3 &  $-$0.67 $\pm$  0.03 \\ 
79 & 16 19 24.24 $\pm$  0.18 & 55 50 52.8 $\pm$   1.6 &      2.18 $\pm$  0.21 &    0.79 $\pm$  0.10 &      19 &   26.8 $\pm$   3.7 & $>-$1.16 \hfill \quad & \\ 
80 & 16 21 13.68 $\pm$  0.10 & 55 23 42.0 $\pm$   0.9 &     58.11 $\pm$  1.49 &    2.62 $\pm$  0.12 &      50 &   4.8  $\pm$   0.3 &  $-$0.94 $\pm$  0.03 \\ 
81 & 16 21 18.72 $\pm$  0.28 & 55 38 27.6 $\pm$   2.3 &      7.97 $\pm$  0.40 &    0.74 $\pm$  0.10 &       7 &   8.5  $\pm$   1.2 &  $-$0.98 $\pm$  0.08 \\ 
82 & 16 21 45.36 $\pm$  0.10 & 55 49 37.2 $\pm$   1.0 &     31.10 $\pm$  1.01 &    2.05 $\pm$  0.11 &      60 &   6.9  $\pm$   0.4 &  $-$0.62 $\pm$  0.04 \\ 
83 & 16 22 08.64 $\pm$  0.14 & 55 24 28.8 $\pm$   1.6 &     14.23 $\pm$  2.15 &    4.55 $\pm$  0.15 &       6 &  27.7  $\pm$   1.2 &  $-$1.57 $\pm$  0.03 \\ 
\enddata
\end{deluxetable}

\begin{deluxetable}{lcccc}
\tablecolumns{3}
\tablewidth{0pc} 
\tablecaption{ Polarized source counts \label{polcounts-tab} }
\tablehead{
\colhead{$p$} & \colhead{\ \ \ \ \ \ \ \ \ } & \colhead{$N$} & \colhead{\ \ \ \ \ } & \colhead{$p^{2.5} dN/dp$}\\ 
\colhead{(mJy)} & \colhead{\ }  & \colhead{\ } & \colhead{\ } & \colhead{(Jy$^{1.5}$ sr$^{-1}$)} \\
}
\startdata
\phantom{0}0.71   &   &           29   &  & 1.38 $\pm$ 0.26    \\
\phantom{0}1.42   &   &           20   &  & 1.26 $\pm$ 0.28    \\
\phantom{0}2.50   &   &           11   &  & 1.10 $\pm$ 0.33    \\
\phantom{0}4.61   &   & \phantom{0}5   &  & 1.01 $\pm$ 0.45    \\
\phantom{0}9.94   &   & \phantom{0}2   &  & 1.46 $\pm$ 1.03    \\
20.4    &   & \phantom{0}1   &  & 2.32 $\pm$ 2.32    \\
\enddata
\end{deluxetable}

\begin{deluxetable}{ccccc}
\tablecolumns{5}
\tablewidth{0pc} 
\tablecaption{ Radio sources in the IRAC color-color diagram \label{table_irac_colcol} }
\tablehead{
\colhead{Region} & \multicolumn{2}{c}{Polarization detected} & \multicolumn{2}{c}{Other sources}\\ 
\colhead{} & \colhead{Number}  & \colhead{\%} & \colhead{Number}  & \colhead{\%}
 \\
}
\startdata
 1                       &\phantom{0}8    &  20$\pm$\phantom{0}7     &  32 &   19$\pm$3 \\
 2                       &\phantom{0}0    &  0  &  25 &   15$\pm$3 \\
 3a\tablenotemark{a}     &  21    &  51$\pm$11    &  80 &   47$\pm$5 \\
 3b\tablenotemark{b}     &  11    &  27$\pm$\phantom{0}8     &  24 &   14$\pm$3 \\
 4                       &\phantom{0}1    &\phantom{0}2$\pm$\phantom{0}2     & \phantom{0}9 &  \phantom{0}5$\pm$2 \\
\enddata
\tablenotetext{a}{Region 3 blue clump, selected by $\log(S_{8.0})/\log(S_{4.5})\leqq -0.5$}
\tablenotetext{b}{Region 3 PAH, selected by $\log(S_{8.0})/\log(S_{4.5}) > -0.5$}
\end{deluxetable}

\end{document}